\documentclass{aa}
\usepackage{epsfig}
\begin{document}
\newcommand{\bea}{\begin{eqnarray}}    
\newcommand{\eea}{\end{eqnarray}}      
\newcommand{\be}{\begin{equation}}
\newcommand{\ee}{\end{equation}}
\newcommand{\bef}{\begin{figue}}
\newcommand{\eef}{\end{figure}}
\newcommand{\etal}{et al.}
\newcommand{\kms}{\,{\rm km}\;{\rm s}^{-1}}
\newcommand{\hubunits}{\,\kms\;{\rm Mpc}^{-1}}
\newcommand{\hmpc}{\,h^{-1}\;{\rm Mpc}}
\newcommand{\hkpc}{\,h^{-1}\;{\rm kpc}}
\newcommand{\msun}{M_\odot}
\newcommand{\K}{\,{\rm K}}
\newcommand{\cm}{{\rm cm}}
\newcommand{\cd}{{\langle n(r) \rangle_p}}
\newcommand{\Mpc}{{\rm Mpc}}
\newcommand{\kpc}{{\rm kpc}}
\newcommand{\xir}{{\xi(r)}}
\newcommand{\xrp}{{\xi(r_p,\pi)}}
\newcommand{\xsirpi}{{\xi(r_p,\pi)}}
\newcommand{\wrp}{{w_p(r_p)}}
\newcommand{\gr}{{g-r}}
\newcommand{\Navg}{N_{\rm avg}}
\newcommand{\Mmin}{M_{\rm min}}
\newcommand{\fiso}{f_{\rm iso}}
\newcommand{\Mr}{M_r}
\newcommand{\rp}{r_p}
\newcommand{\zmax}{z_{\rm max}}
\newcommand{\zmin}{z_{\rm min}}

\def\eg{{e.g.}}
\def\ie{{i.e.}}
\def\spose#1{\hbox to 0pt{#1\hss}}
\def\ltapprox{\mathrel{\spose{\lower 3pt\hbox{$\mathchar"218$}}
\raise 2.0pt\hbox{$\mathchar"13C$}}}
\def\gtapprox{\mathrel{\spose{\lower 3pt\hbox{$\mathchar"218$}}
\raise 2.0pt\hbox{$\mathchar"13E$}}}
\def\inapprox{\mathrel{\spose{\lower 3pt\hbox{$\mathchar"218$}}
\raise 2.0pt\hbox{$\mathchar"232$}}}

\title{Basic properties of galaxy clustering in the light
of recent results from the Sloan Digital Sky Survey} 

\subtitle{}

\author{Michael Joyce\inst{1}, Francesco Sylos Labini \inst{2,3},
Andrea Gabrielli \inst{3,4}, Marco Montuori \inst{3,4} and Luciano
Pietronero \inst{3,4}}

\titlerunning{Galaxy clustering from the SDSS} 
\authorrunning{Joyce et al.} 

\institute{Laboratoire de Physique Nucl\'eaire et des Hautes
Energies, Universit\'e de VI, 4, Place Jussieu, Tour 33 - Rez de
Chaus\'ee, 75252 Paris Cedex 05, France
\and ``Enrico Fermi Center'', Via Panisperna 89 A, Compendio del Viminale, 00184 Rome, Italy
\and ``Istituto dei Sistemi Complessi'' CNR, Via dei Taurini 19, 00185 Rome, Italy
\and 
Statistical Mechanics and Complexity Center ---
Istituto Nazionale Fisica della Materia, Unit\'a di Roma 1, and
Dipartimento di Fisica, Universit\'a di Roma ``La Sapienza'',
P.le A. Moro 2, 00185 Rome, Italy}

\date{Received / Accepted}

\abstract{We discuss some of the basic implications of recent results on galaxy
correlations published by the SDSS collaboration.  In particular we
focus on the evidence which has been recently presented for the scale
and nature of the transition to homogeneity in the galaxy
distribution, and results which describe the dependence of clustering
on luminosity. The two questions are in fact strictly entangled, as
the stability of the measure of the amplitude of the correlation
function depends on the scale at which the mean density becomes well
defined. We note that the recent results which indicate the
convergence to well defined homogeneity in a volume equivalent to that
of a sphere of radius 70 Mpc/h, place in doubt previous detections of
``luminosity bias'' from measures of the amplitude of the correlation
function.  We emphasize that the way to resolve these issues is to
first use, in volume limited samples corresponding to different ranges
of luminosity, the unnormalized two point statistics to establish the
scale (and value) at which the mean density becomes well defined.  We
note also that the recent SDSS results for these statistics are in
good agreement with those obtained by us through analyses of many
previous samples, confirming in particular that the galaxy
distribution is well described by a fractal dimension $D \approx 2$ up
to a scale of at least $20$ Mpc/h.  We discuss critically the
agreement of this new data with current theoretical models.
 }

\maketitle

\section{Introduction}

The most striking feature of the large scale distribution of galaxies
is their organization in structures as clusters, super-clusters and
filaments around large volumes of space empty of visible matter, the
voids. Since the first observations of the three-dimensional
distribution of galaxies (Kirshner et al. 1983) evidence has
accumulated for the existence of such large agglomerations of
matter. In the eighties the ``Great Wall'', a giant filament with an
extension of about $200$ Mpc/h connecting many galaxy groups and
clusters, was discovered (De Lapparent, Huchra \& Geller 1989).
Currently the largest known structure in the local universe is
the recently discovered ``Sloan Great Wall'' (Gott et al. 2005),
roughly twice longer than the Great Wall. Its discovery has been
possible with new data provided by the Sloan Digital Sky Survey (SDSS
--- York et al. 2000), one of the most ambitious observational
programs ever undertaken in astronomy. It will measure about 1 million
redshifts, giving a complete mapping of the local universe up to a
depth of several hundreds of Mpc.  In this paper we discuss the
results, and implications, of some recently published studies of the
correlation properties of the galaxy distribution based on new data
from the partially completed SDSS.

\section{Characterization of galaxy correlations} 

An accurate statistical characterization of galaxy structures is
evidently a key element for any physical theory explaining their
origin.  The earliest observational studies, from angular catalogs,
produced the primary result (Totsuji \& Kihara 1969) that the {\it reduced
two-point correlation function} $\xi(r) \equiv \frac{\langle n(r)n(0)
\rangle}{\langle n \rangle^2}-1$ (where $n$ is the density of points)
is well approximated, in the range of scales from about $0.1$ Mpc/h to
$10$ Mpc/h, by a simple power-law\footnote{The subscript $E$ indicates 
the statistical estimator of $\xi(r)$ in a given sample}: 
\be
\label{xi_stand}
\xi_E(r)\approx \left(\frac{r}{r_0}\right)^{-\gamma} \ee with $\gamma
\approx 1.8$ and $r_0 \approx 4.7$ Mpc/$h$.  This result was
subsequently confirmed by numerous other authors in different redshift
surveys (e.g., Davis \& Peebles 1983; see Peebles 2001 for a recent
review). However, while $\xi_E(r)$ shows consistently a simple
power-law behavior characterized by this exponent, there is very
considerable variation among samples, with different depths and
luminosity cuts, in the measured {\it amplitude} of $\xi_E(r)$.  This
variation is usually ascribed {\it a posteriori} to an intrinsic
difference in the correlation properties of galaxies of different
luminosity --- ``luminosity bias'' (see e.g., Davis et al. 1998):
brighter galaxies present larger values of $r_0$.  Theoretically it is
interpreted as a real physical phenomenon, as a manifestation of
``biasing'' (Kaiser 1984), or, more recently, in terms of ``halo
models'' (see e.g. Cooray \& Sheth 2002).

Such a variation of the amplitude of the measured correlation function
may, however, be explained, entirely or partially, as a finite-size
effect i.e. as an artifact of statistical analysis in finite samples.
The explanation is as follows (see Gabrielli et al. 2004 --- hereafter
GSLJP --- for a detailed discussion, and original references). The
reduced correlation function $\xi(r)$ can be written as
$\xi(r)=\frac{\langle n(r)\rangle_p}{\langle n \rangle}-1$, where
${\langle n(r)\rangle}_p$ is the {\it conditional density} of points
i.e. the mean density of points in a spherical shell of radius $r$
centered on a galaxy (the subscript $p$ indicates the condition that
the density is measured from an occupied point).  The latter is
generally a very stable local quantity, the reliable estimation of
which at a given scale $r$ requires only a sample large enough to
allow a reasonable number of independent estimates of the density in a
shell. The mean density ${\langle n \rangle}$, on the other hand, is a
global quantity. The size of a sample in which it is estimated
reliably is not known {\it a priori}, but depends on the properties of
the underlying distribution.  Specifically the sample must be large
enough so that the mean density estimated in it has a sufficiently
small fluctuation with respect to the true asymptotic average
density. If $\xi^{(1)}_E(r)$ and $\xi^{(2)}_E(r)$ are estimates of
$\xi(r)$ in two samples with, respectively, mean densities
$\bar{n}^{(1)}_S$ and $\bar{n}^{(2)}_S$, one has
\begin{equation}
\xi^{(2)}_E(r) \approx \frac{\bar{n}^{(1)}_S}{\bar{n}^{(2)}_S} \xi^{(1)}_E(r)
\end{equation}
in the regime where $\xi^{(1)}_E(r), \xi^{(2)}_E(r) \gg 1$.  A linear
amplification of $\xi_E(r)$ in more luminous samples will thus be
observed if, for example, the fainter galaxy samples probe volumes
smaller than those at which the mean density becomes well defined in
the underlying distribution. There is in fact such a systematic effect
in observations, as fainter galaxies can only be seen at smaller
distances.

\section{Finite size effects in the measurements of galaxy correlations} 

It has been pointed out by Pietronero (1987) that, when analyzing a
point distribution which, like the galaxy distribution, is
characterized by large fluctuations, one should, in fact, first
establish the existence of a well defined mean density (and ultimately
the scale at which it becomes well defined and independent of the
sample size, if it does) before a statistic like $\xi(r)$, which
measures fluctuations with respect to such a mean density, is
employed. Further the existence of power-law correlations, which are
clearly present in the galaxy distribution, is typical of fractal
distributions, which are asymptotically empty.  In such distributions
the mean density is always strongly sample dependent, with an average
value decreasing as a function of sample size. Given the observation
of such correlations in the system, and the instability of the
amplitude of the correlation function $\xi_E(r)$ estimated in
different samples, special care should be taken in establishing first
the scale (if any) at which homogeneity becomes a good approximation.
The simplest way to do this is in fact to measure the conditional
density ${\langle n(r)\rangle}_p$, or, alternatively, ${\langle
n^*(r)\rangle}_p$, the {\it integrated conditional density}, where
$n^*(r)$ represents the density in a {\it sphere} of radius $r$ about
an occupied point. These quantities are generally well defined, and
give a characterization of the two-point correlation properties of the
distribution, irrespective of whether the underlying distribution has
a well defined mean density or not.  A simple power law behavior $\cd
= B r^{-\gamma}$ is characteristic of scale-invariant fractal
distributions, with the exponent $\gamma<3$ giving the {\it fractal
dimension} through $D=3-\gamma$. The pre-factor $B$ is, in this case,
simply related to the lower cut-off of the distribution (GSLJP)
\footnote{We remark that in nature any fractal distribution has a
lower and an upper cut-off between which the characteristic scaling
proprieties are manifested: the infinite volume limit is, of course,
simply a useful mathematical idealization. In particular the
hypothesis that the galaxy distribution manifests fractal scaling
always refers implicitly to some finite range of (observable)
scales.}.  If the distribution has a well defined mean density, one
has, asymptotically, ${\langle n(r)\rangle}_p={\langle
n^*(r)\rangle}_p=const. >0 $ (i.e., $D=3$ in the previous
formula). Measurement of these quantities can thus both characterize
(i) the regime of strong clustering and (ii) the scale and nature of a
transition to homogeneity.  Only once the existence of an average
density within the sample size is established in this manner does it
make sense to use $\xi(r)$. And, in that case, the analysis with
${\langle n(r)\rangle}_p$ can determine the minimal size of samples
required to make the use of this quantity meaningful.

In studies of galaxy data from numerous different surveys using the
conditional density (see GSLJP and references therein, in particular
Joyce, Montuori \& Sylos Labini 1999) it has been found that, at
scales up to roughly $20$ Mpc/h, where the statistics are very robust
(i.e. where a reasonable number of independent spherical shells can be
fully inscribed in the sample volume), the galaxy distribution is well
described by a simple fractal scaling with $D \approx 2$. At larger
scales (up to $\sim 100$ Mpc/h, or even greater) it has been argued on
the basis of these analyses that there is weaker statistical evidence
for the continuation of a fractal scaling, and no clear evidence for
homogeneity.  These analyses not only place in question the physical
meaning, of the {\it amplitude} of $\xi(r)$, as discussed above, but
also produce a different value of the {\it exponent} characterizing
galaxy clustering at small scales\footnote{Note that we are discussing
here the results of measures of $\xi(r)$ and ${\langle n(r)\rangle}_p$
{\it in redshift space}. Such studies (e.g. Park et al. 1994, Benoist
et al. 1996) have consistently reported a dimension $D=1.2 -
1.5$. Note also that finite size effects perturb the estimations of
the correlation exponent differently for the correlation function and
the power spectrum (see Sylos Labini \& Amendola 1996 for more
details).}. Indeed, as $\xi(r) \approx {\langle n(r)\rangle}_p$ at the
scale where $\xi(r) > 1$, the two statistics should agree at these
scales. The discrepancy can be explained (see GSLJP) as a result of
the way in which the dimension $D$ is normally estimated from
$\xi_E(r)$: it is determined by fitting a power-law in a log-log plot
around the scale $r_0$ defined by $\xi_E(r_0)=1$. If there is a strong
break from simple power-law behavior in $\xi(r)$ around this scale the
estimated dimension $D'=3-\gamma'$ is systematically smaller than
$D=3-\gamma$ (e.g. with $\xi(r)=(r_0/r)^\gamma e^{-r/r_0}$ one has
$\gamma'=-d(\ln
\xi)/d(\ln r)|_{r=r_0}=\gamma+1$).
For example one may note that Hawkins et al. (2003) by fitting
$\xi_E(r)$ in the regime $r<r_0$ find that $\gamma\approx 0.75$, while
at scales of order $r_0$ they measure $\gamma\approx 1.6$.  On the
other hand Tikhonov et al. (2003) measured a value of about
$\gamma=0.6\div 0.8$ by studying the conditional density in different
galaxy samples; in addition they found evidences for a crossover to
homogeneity for scales large than 100 Mpc/h.  Similar results were
found by Baryshev \& Bukhmastova (2004) in some early SDSS samples, by
considering the two-point conditional density.

\section{New SDSS results on galaxy correlations} 

In HEB3GS the integral conditional density ${\langle
n^*(r)\rangle}_p$ has been estimated in the ``luminous red galaxy''
(LRG) sample of the SDSS survey. It is an approximately
volume limited (VL) sample 
with very precise photometric calibration, and is by far the
largest such sample ever considered for such an analysis.  Its size
allows the robust estimation of this statistic up to a scale of 
order $100$ Mpc/h, as at this scale the sample contains a considerable
number of independent (non-overlapping) spheres centered on
galaxies. The results can be summarized as follows: (i) A simple
power-law scaling corresponding to a fractal dimension $D=2$
gives a very good fit to the data up to at least $20$ Mpc/h, over 
approximately a decade in scale; (ii) at 
larger scales ${\langle n^*(r)\rangle}_p$ continues to decrease,
but less rapidly, until about $\sim 70$ Mpc/h, above which
it flattens up to the largest scale probed by the sample 
($100$ Mpc/h).
The transition between the two regimes is slow, in
the sense that the integrated conditional density at $\sim 20$ Mpc/h is 
about twice the asymptotic mean density.

Let us consider the conclusions which can be drawn from this analysis
in view of the discussion in the previous section.  Firstly, the
results are highly consistent with the claim that galaxy correlations,
up to approximately $20$ Mpc/h, are well characterized by a simple
power-law scaling corresponding to a fractal dimension $D\approx 2$.
The LRG sample probes the scaling of ${\langle n^*(r)\rangle}_p$ only
from about $3$ Mpc/h, and evidently only for the brighter galaxies
represented by this sample. The same exponent has, however, been
observed to describe well the behavior of ${\langle n^*(r)\rangle}_p$
by our analysis of many precedent galaxy samples over a large range of
luminosities, and down to scales an order of magnitude smaller (see
GJSLP and references therein).

Secondly, the analysis provides good statistical evidence for
homogeneity at larger scales, with a flattening of ${\langle
n^*(r)\rangle}_p$ apparent above $\lambda_0 \approx 70$ Mpc/h.  Future
data, and in particular the even larger forthcoming samples from SDSS,
will confirm or refute this very important result, extending the range
of scale between the detected $\lambda_0$ and the sample
size\footnote{We remark, in particular, that in HEB3GS the density in
each sphere is not actually calculated by dividing the number of
points by its volume. Instead the division is by the number of points
in the same volume in an artificial catalog, which is constructed to
correct for effects leading to incompleteness of the sample. This
actually involves building in the assumption of homogeneity at large
scales. It is stated that the sample is sufficiently complete that
this is ``essentially equivalent'' to dividing by the volume.  The
dependence of the results on this difference should however be
quantified carefully and controlled for in future analyses.}.  If one
takes this scale to be a reliable determination of the scale of
homogeneity, it implies that $\xi(r)$ is a well defined statistical
quantity.  Further one can infer the size of samples in which one can
usefully employ this statistic to characterize the statistical
properties of the galaxy distribution. Here ``usefully'' means that
the finite size effects which can systematically offset the amplitude
of the estimator $\xi_E(r)$ from that of $\xi(r)$ are under control,
so that a physical significance can be attributed to this amplitude
(to a degree of accuracy which can be determined). For this to be the
case one requires that the sample be large enough so that (i) the
sample mean density approximates sufficiently well to the true
asymptotic mean density, and (ii) the sample to sample fluctuations
(i.e. variance) in the density are sufficiently small. The requirement
on the sample size imposed by (i) is determined by comparing the
conditional density at the depth of the sample with the measured
asymptotic density, and the results of HEB3GS  indicate that a sample
of depth greater than $\approx 70$ Mpc/h is sufficient to make such
systematic effects small.  To determine the constraint imposed by (ii)
one needs a measure of fluctuations in the density in such a sample
about the (now well defined) mean density.  HEB3GS provides, beyond
the measure of the conditional density (which gives information about
the average and not the fluctuations), one measure of such
fluctuations in the LRG sample: for $0.2 < z < 0.35$ in ten disjoint
regions with volumes $V \approx 2.2 \times 10^7$ (Mpc/h)$^3$
field-to-field fluctuations are measured to be $\frac{\sqrt{\langle
\Delta n^2\rangle} } {\langle n \rangle} \approx 0.07$, after
subtraction of a Poisson noise term.  Such fluctuations, in general,
depend not only on the volume of the region, but also on its
geometry. However, if the sample is sufficiently large to include
several spheres of radius of order $\lambda_0$, we can suppose such an
equivalence. Using the estimations of HEB3GS  we would thus expect
that, to reduce the finite size fluctuations in the amplitude of
$\xi_E(r)$ to of order ten percent, we need (i) samples including
complete spheres of radius $\lambda_0\approx 70$ Mpc/h, and (ii) an
equivalent depth at least $R_e \approx \left(\frac{4\pi}{\Omega}
\right)^{1/3} 175$ Mpc/h (using $\frac{\Omega}{3}R_e^3=V$).

Let us consider how these values for $\lambda_0$ and $R_e$ compare
with those of redshift surveys prior to SDSS: in the 2dF survey
(e.g. Peacock et al. 2001), which has $\Omega \approx 0.05$ sr, the largest
enclosed sphere in a VL sample is of radius $\sim 30$ Mpc/h and the
greatest depth of such a sample is considerably smaller than $R_e
\approx 450$ Mpc/h; in the ESP catalog (Guzzo et al. 1999) the
radius of the largest enclosed sphere is $\sim 20$ Mpc/h, while the
largest VL sample has a volume approximately one hundred times smaller
than that corresponding in the LRG sample to fluctuations of order
$10\%$ in the density. The determinations of HEB3GS imply that these,
and indeed all, previous measures of $\xi(r)$ in redshift catalogs
prior to SDSS are expected to be severely affected by the finite size
effects we have discussed.

\section{Implications with respect to other results from the SDSS 
collaboration} 

In another recent paper (Zehavi et al.2004A) the SDSS collaboration
reports a study of the correlation properties of a sample of 200,000
galaxies, covering a solid angle of about $4\pi/16$, with the standard
method of $\xi(r)$ estimation.  The paper focuses on the dependence of
this quantity on galaxy properties, finding in particular that the
amplitude of the measured two-point correlation function rises
continuously with absolute magnitude from $\Mr \approx -17.5$ to $\Mr
\approx -22.5$, with the most rapid increase occurring above the
characteristic luminosity $L_*$ ($\Mr \approx -20.5$). The scale $r_0$
(determined by $\xi_E(r)=1$) varies in the range $[2.7,10.0]$ Mpc/h
\footnote{The paper actually follows what has become a standard
practice in the last few years: results are given only for the
``projected real space'' correlation function, rather than the
redshift space correlation function, which we have considered here.
This makes it much more difficult to compare the results for the
estimated dimension, and so we will not focus on this point here.}.

As we have discussed above, the results of HEB3GS indicate that one
expects finite size effects in the amplitude of $\xi_E(r)$ to be
reduced to of order ten percent under two conditions: the samples (i)
enclose fully a spheres of radius $\approx 70$ Mpc/h, and (ii) have at
least a depth $R_e \approx \left(\frac{4\pi}{\Omega} \right)^{1/3} 170
$Mpc/h.  These criteria are in fact satisfied only by the largest VL
samples considered in Zehavi et al. (2004A), and so there is a
priori evidence that such finite size effects, and not a real physical
effect, may be, wholly or partially, responsible for the measured
variation in $r_0$.  Zehavi et al.(2004A) do actually consider
indirectly this possibility, and report a test of this hypothesis in
which the correlation functions are measured in adjacent luminosity
samples cut to cover the same spatial volumes. Of the four pairs of
samples there is reasonable stability in only one case, in the three
others the result clearly favors the hypothesis of volume
dependence. The authors argue that their results indicate the opposite
conclusion, and ascribe the observed variation between the different
volumes a posteriori to ``anomalous'' fluctuations of the density in
specific samples due to the presence of the ``Sloan Great
Wall''. Rather than being anomalous, such fluctuations are simply
indicative of the intrinsic correlations of the galaxy distribution.

In a subsequent paper by the SDSS collaboration (Zehavi et al. 2004B)
a standard $\xi(r)$ analysis is applied to the same LRG sample of
galaxies considered in HEB3GS. In this case HEB3GS provides a
justification for the use of this statistic, as it has shown that the
sample does contain many independent spheres with radius larger than
that at which the conditional density flattens.  What is found in
Zehavi et al.(2004B) is that $\xi_E(r)$ shows a considerable degree of
stability in different samples. In particular the full correlation
function of the LRG galaxies in redshift space is, within the
estimated error bars, in very good agreement with that of the most
luminous galaxies (i.e. largest volume samples) measured in Zehavi et
al. (2004 A).  This appears to give a quite robust determination of
the scale $r_0 \approx 13$ Mpc/h
\footnote{This value (in redshift space) can be inferred from the
data given in Table 2 of Zehavi et al. (2004B).}, the stability of
which it will be interesting to see confirmed by forthcoming larger
samples of SDSS.  On breaking the LRG sample into three sub-samples by
luminosity, however, Zehavi et al.(2004B) finds some variation in the
amplitude of $\xi_E(r)$, which is attributed again to ``luminosity
bias''. The variation is however at a level of only $20 \%$ in the
amplitude, compared to a factor of almost ten between the brightest
and faintest sample in Zehavi et al.(2004A).  This is at a comparable
level to the fluctuations in the mean density found at the scale of
the sample in HEB3GS and so this ``detection'' of luminosity bias
appears highly questionable. Indeed Zehavi et al. (2004B) measure
$\xi_E(r)$ in three different subsamples of the LRG sample,
corresponding to three redshift bins, and find variations in
amplitude {\it at the same level}.  In this case they conclude that
this variation is ``likely to reflect large-scale structure
variations'' i.e. precisely the finite size effects we are discussing.

To clearly determine whether the observed variation in amplitude is a
manifestation of a real difference between galaxies of different
luminosity, or a volume effect, for which we have argued there is a
better prima facie case, requires, we believe, a much more systematic
statistical analysis.  The simplest way to perform such an analysis is
via the conditional density ${\langle n(r)\rangle}_p$ (or
alternatively via ${\langle n^*(r)\rangle}_p$), performed on galaxies
with different properties (and, in particular, luminosities) in VL
samples.  By doing so, one obtains both a characterization of the
two-point correlation properties (and in particular exponent or
exponents) characterizing the clustering in the strongly clustered
regime, and determines the scale at which homogeneity is
established. A fundamental question is then whether galaxies of
different luminosity show the same behavior of $\cd$, and, more
specifically, whether they are characterized by the same scaling
dimension in the regime of power-law correlations, and the same
homogeneity scale $\lambda_0$.  The considerable amount of data we
have analyzed in the past (for discussion and references, see GSLJP)
indicated that $\cd$ has approximately the same simple power-law
behavior $D\approx 2$ in the ranges of scales probed, with only a very
weak dependence on luminosity.  Once such a systematic analysis has
been performed, and the homogeneity scale (or scales) established, a
$\xi(r)$ analysis can be used to characterize correlation in the
regime of small amplitude fluctuations.  Note that a linear
amplification of $\xi(r)$, with no distortion in the range $\xi(r)
>1$, as a function of luminosity would correspond in the $\cd$
analysis to an observed dependence of the homogeneity scale,
$\lambda_0$, on galaxy luminosity.

\section{Theoretical Implications \& Open Questions} 

We conclude with some remarks on other related observational
and theoretical questions about galaxy correlations.

We have underlined that the results of HEB3GS for the conditional
density confirm the characterization, to a very good approximation, of
the galaxy distribution by a single exponent $D \approx 2$ up to $20$
Mpc/h. The most straightforward interpretation of such an observation
is that it indicates a scale-invariant fractal behavior over the
corresponding range of scale. However, other explanations are possible
which invoke an underlying distribution which is not fractal (see
GSLJP, and references therein). One such possibility is a distribution
of spherically symmetric density profiles described by a rapidly
decaying power-law about their centers.  Indeed this corresponds to
what is observed in cosmological N body simulations of dark matter and
used in phenomenological models (``halo models'', see Cooray \& Sheth
2002) for the formation of structure by gravity. One can perform
further statistical tests to differentiate the two qualitatively
different distributions, notably by more direct determinations of the
dimension by box-counting methods, or of the behavior of the
conditional variance as a function of scale (Bottaccio et al. 2004;
GSLJP).  Results obtained by some of us (Coleman \& Pietronero 1992;
Sylos Labini, Montuori \& Pietronero 1998) support the fractal
interpretation, but with a limited statistical confidence. With the
data now becoming available from SDSS it should be possible to
determine the answer to this basic question about the nature of the
galaxy distribution more definitively.

Beyond this specific basic question about the nature of the galaxy
distribution at scales well below the homogeneity scale, i.e. in the
regime of strong fluctuations, a more general question concerns the
agreement between current cosmological models and the data on the
galaxy distribution over all scales probed by current data.  HEB3GS
states that the correlation properties of LRG galaxies measured are
compatible with those predicted in a standard $\Lambda$CDM model, with
a ``bias factor'' of order 2. In this paper the evidence for this
conclusion comes solely from the comparison, which gives the quoted
bias factor, of the amplitude of the measured fluctuations in the
volume equivalent to a sphere of radius $175$ Mpc/h with the
prediction of this theoretical model, rather than from a comparison of
the model with the data over a range of scales.  In Fig.\ref{fig1} we
show on a single plot the results of HEB3GS and the integrated
conditional density calculated in a large $\Lambda$CDM simulation of
dark matter. The theoretical ${\langle n^*(r)\rangle}_p$ of the dark
matter shows an approximately power-law behavior, characterized by an
exponent $\gamma \approx 1.8$, followed by a {\it rapid} cross-over to
homogeneity at a scale $\lambda_0 \approx 10$ Mpc/h. Also shown are
the results from the analysis of the CfA2 survey reported in Joyce,
Montuori \& Sylos Labini (1999).

\begin{figure}
\begin{center}
\includegraphics*[angle=0, width=0.5\textwidth]{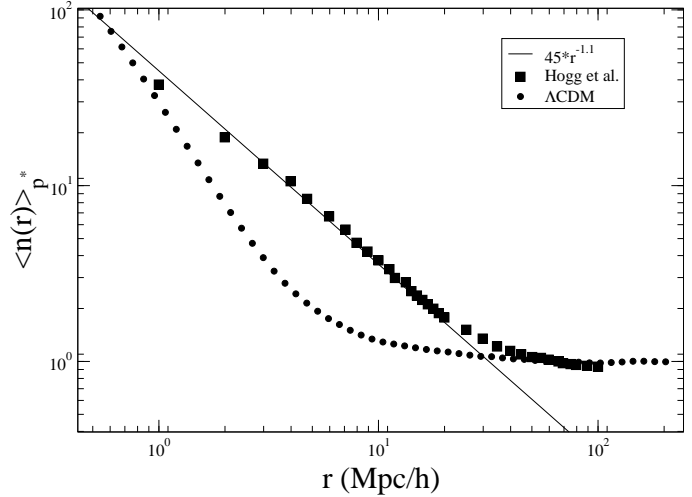}
\end{center}
\caption{Behavior of $\cd$ for ``CDM particles'' in a
$\Lambda$CDM simulation run by the Virgo consortium (Jenkins et
al. 1998) together with the determination by HEB3GS and those in
several samples of the CfA2 catalog reported in Joyce, Montuori \&
Sylos Labini (1999).  Note that (i) the power-law index in the
simulated data is quite different to that in the observations ($\gamma
\approx 1.8$ instead of $\gamma \approx 1$ --- Baertschiger, Joyce \&
Sylos Labini 2002); (ii) the homogeneity scale detected in the LRG
sample, $\lambda_0 \approx 70$Mpc/h, is very significantly larger than
that in the simulated data ($\lambda_0 \approx 10$ Mpc/h). See text
for discussion.
} \label{fig1}
\end{figure}

What is implicitly assumed by HEB3GS is that the formation of LRGs
can be described as a sampling procedure on the underlying dark
matter, which leads, on the one hand, to a linear amplification of the
fluctuations by a factor of 2 at very large scales, while also
producing the correlation properties observed in the LRGs from the
very different ones of the underlying dark matter field. So-called
``halo models'' (see e.g.  Cooray \& Sheth 2002) currently provide the
framework to describe such a sampling: the distribution of dark matter
is described as a set of halos, to which formation probabilities for
the different kinds of objects are ascribed {\it a posteriori} to
produce the observed correlation properties.  Given that the
properties of the two distributions are so different, we believe it is
fair to say that the statement that these observations accord with the
theory is very weak: any set of observations would probably accord
with theory given this criteria.

Given that the notion of ``bias'', which has its origin in the
measurements of variations in the amplitude in $\xi_E(r)$, is so
central in these theoretical constructions, it is evidently crucial to
correctly understand the role of finite size effects in determinations
of amplitudes of correlation functions, using the method we have
outlined.  Only once this is done can the theoretical problem in the
regime of strong fluctuations be properly addressed.  The simplicity
of the correlation properties of visible matter over a range of two
decades in scale --- well described by a simple power-law scaling from
$\sim 0.1$ Mpc/h to approximately $20$ Mpc/h --- cannot but suggest
that the relegation by current theories of visible matter to an
epiphenomenon of the underlying dark matter may be mistaken.

On larger scales, where fluctuations become small, the comparison
between theory and data becomes in principle much cleaner, as the
theoretical description of this regime becomes simpler. Measurement of
the signal, which is of very low amplitude, becomes, however, much
more difficult, and indeed it is only with data from SDSS now emerging
that it is becoming feasible to make such comparisons. Indeed a very
recent paper published by the SDSS collaboration
(Eisenstein et al. 2005) give an analysis of galaxy correlations in this
regime and claims a statistically significant detection of a
``bump-like'' feature in the real space correlation function $\xi(r)$
at the location predicted by the same $\Lambda$CDM model considered in
HEB3GS (for a discussion of these real space properties of standard
models, see GSLJP, Chap. 6).  Evidently such findings, if confirmed by
other galaxy data, provide dramatic evidence in favor of the standard
model of structure formation. We note (Gabrielli Joyce \& Sylos Labini
2002; GJSLP) that a very characteristic prediction in this regime of
standard models is the existence of a negative power-law tail with
$\xi(r) \sim -1/r^4$ at the very largest scales which the SDSS survey
is now beginning to probe. Such a behavior is the direct signal of
the ``primordial'' form of the fluctuations in standard cosmological
models, with a power spectrum which is linear in $k$ at small
$k$. Interestingly the simple behavior of the power spectrum is
destroyed by the ``sampling'' of the CDM field (i.e. by a prescription
for how to infer galaxy correlations from those of dark matter), while
that of the real space correlation function remains intact (Durrer et
al. 2003; GJSLP). This behavior can be understood simply in terms of
the very specific ``super-homogeneous'' properties of the primordial
fluctuation field, which are also encountered (Gabrielli et al. 2003)
in certain systems widely studied in statistical physics.
  
\begin{acknowledgements}
We warmly thank Daniel Eisenstein and David Hogg for
very interesting discussions on some of the issues discussed here.
Moreover, it is a real pleasure to acknowledge collaborations and
discussions with Yurij Baryhsev, Thierry Baertschiger, Helene Di
Nella-Courtois, Ruth Durrer, Bill Saslaw, and Pekka Teerikorpi.
\end{acknowledgements}

{}

\end{document}